\documentclass[9pt,twocolumn,twoside]{opticajnl}
\journal{opticajournal} % use for journal or Optica Open submissions

\setboolean{shortarticle}{true}
\usepackage{lineno}
\usepackage{mathrsfs}
\usepackage{xcolor}
\usepackage{graphicx}
\usepackage{bm}% bold math
\title{Phase-dependent THz yield in two-color laser plasmas: nonlinear propagation and Coulomb-modified recollision}

\author[1,2]{Yange Chen}
\author[3,*]{Ahai Chen}
\author[3]{Jiahui Guo}
\author[2]{Tianmin Yan}
\author[4]{Yizhu Zhang}
\author[3,5,1,*]{Yuhai Jiang}
\affil[1]{{Shanghai Advanced Research Institute}, {Chinese Academy of Sciences}, {{Shanghai 201210}, {China}}}
\affil[2]{{University of Chinese Academy of Sciences}, {{Beijing 100049}, {China}}}
\affil[3]{{Center for Transformative Science, School of Physical Science and Technology}, {ShanghaiTech University}, {{Shanghai 201210}, {China}}}
\affil[4]{School of Electrical Engineering, Tiangong University, Tianjin 300387, China}
\affil[5]{School of Physics, Henan Normal University, Xinxiang 453007, China}
\affil[*]{chenah1@shanghaitech.edu.cn}
\affil[*]{jiangyh3@shanghaitech.edu.cn}
\begin{abstract}
Two-color laser-induced plasmas provide a reliable source of broadband terahertz (THz) radiation and offer several control parameters for optimizing THz yields.
We investigate the phase-dependent THz yield (PTY) driven by  a two-color laser field with parallel linear polarizations,
using a model that couples macroscopic nonlinear propagation and microscopic Coulomb effects. % d 
Comparisons with models accounting for either effect alone show that their coupled effects shift the yield extrema and reshape the phase-dependent response. 
Nonlinear propagation distorts the local laser waveform, while the Coulomb interaction modifies the subsequent electron dynamics, particularly recollision. These changes modify the directional asymmetry of the photoelectron momentum distribution and hence the  THz emission.
Within the studied range, the combined model predicts a larger shift of the optimal phase than either effect alone, together with a reduced phase separation between the yield maximum and its adjacent minimum.
These findings may guide THz optimization with more complex driving fields.
\end{abstract}
\setboolean{displaycopyright}{false} % Do not include copyright or licensing information in submission.

\begin{document}

\maketitle

\section{Introduction} 
Terahertz (THz) radiation bridges the microwave and infrared regions of the electromagnetic spectrum and features low photon energy, the ability to penetrate many nonmetallic materials, and sensitivity to molecular and material resonances. 
THz techniques have found applications in time-domain spectroscopy~\cite{Fischer:2005,Han:2001}, imaging, communication, and security screening.
These applications motivate the development of powerful, tunable THz sources with controllable waveforms.

THz generation methods include optical rectification in nonlinear crystals~\cite{Hebling:2008}, photoconductive antennas~\cite{Berry:2014}, 
electron accelerators~\cite{Gopal:2013}, and laser-induced gas plasmas~\cite{Kim:2008}.
Among these, the two-color laser (TCL)-induced air plasma scheme 
produces high-intensity, broadband THz pulses~\cite{Cook:2000}.
In this scheme, the superposition of the two laser components breaks the symmetry of the laser waveform, giving rise to a transient current that radiates THz waves far more efficiently than that driven by a single-color field~\cite{Kim:2008}.
It also avoids the damage-threshold and material-absorption limits of solid-state emitters.
Moreover, THz generation can be manipulated by parameters
including pump laser beam profile~\cite{Liu:2016}, TCL wavelength~\cite{Koulouklidis:2020}, and particularly the relative phase $\phi$ of the two laser components~\cite{Mi:2022}.
The $\phi$ dependence provides insight into THz generation, particularly for TCL with with parallel linear polarizations~\cite{Chen:2015, Tailliez:2020}.
Scanning $\phi$ modulates the THz yield over the optical period, with its maxima locked to the timing of the attosecond ionization bursts~\cite{Zhang:2012}.
Moreover, a chirped TCL with a time-varying $\phi$ can generate tunable multicolor and super-continuum THz radiation~\cite{Silaev:2020},
while pulse trains with controlled $\phi$ and carrier-envelope phase can produce narrow-band, frequency-tunable THz radiation~\cite{Zhou:2023}.
In addition, recent reports have shown that the THz spectrum at different frequencies can be modulated by controlling $\phi$~\cite{Zhang:2025}.
scheme~\cite{Babushkin:2022}.

In essence, the $\phi$-dependent THz emission is influenced by the both macroscopic nonlinear propagation and microscopic Coulomb-modified electron dynamics~\cite{Xiao:2021,You:2012}.
Macroscopically,
the interplay of Kerr self-focusing, plasma defocusing, and diffraction~\cite{Couairon:2007} modulates the local TCL amplitudes and $\phi$,
thus affecting the coherent buildup of the THz radiation~\cite{Babushkin:2010}.
For plasma lengths exceeding the dephasing length, longitudinal interference can suppress on-axis emission and produce an on-axis dip in the far-field pattern~\cite{You:2012,Stathopulos:2024}.
These effects are normally addressed by combining nonlinear propagation with a photocurrent model (NP-PC)~\cite{Esarey:1997, Zhang:2022}.
Microscopically,
the parent-ion Coulomb potential modifies the asymptotic photoelectron momentum,
thereby changing the THz polarization and introducing 
$\phi$ dependence features absent from the simple photocurrent picture~\cite{Gan:2024a, Gao:2023}. 
This can be described by models such as classical trajectory Monte Carlo (CTMC)~\cite{Gan:2024a,Gan:2024b, Gao:2023}, time-dependent Schr\"odinger equation simulation~\cite{Babushkin:2022,Chen:2015}, and trajectory-based Coulomb-corrected strong-field approximation~\cite{Zhang:2020b}. 
For parallel linearly polarized TCL, these effects are especially significant on THz yield due to the strong recollision after tunnel ionization~\cite{Chen:2015,Gao:2023}, in contrast to the circularly polarized case, where the Coulomb effect introduces an additional deflection in the ionization direction~\cite{Chen:2026}.

In this work, we address such $\phi$-dependent THz yield (PTY) in a plasma driven by parallel linearly polarized TCL, accounting for both nonlinear propagation and Coulomb effects.
We benchmark the model against experimental data and compare its predictions with conventional models over a range of laser parameters.
These comparisons clarify the roles of the two effects and inform phase optimization for THz generation.

\section{Theoretical Model}
\begin{figure}[t!!]
\centering
\includegraphics[width = 0.940\linewidth]{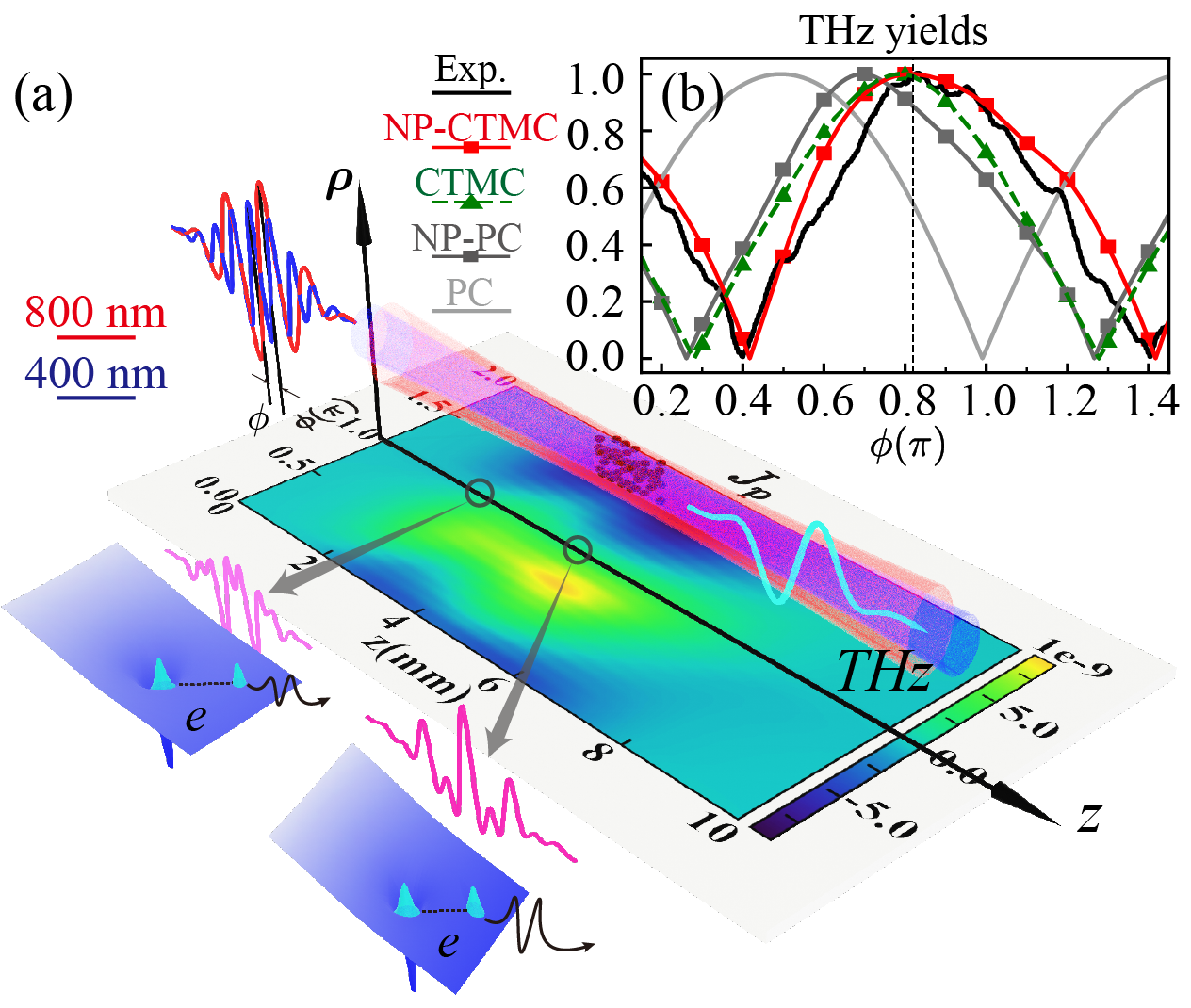}
\caption{
(a) Schematic of NP-CTMC for THz generation with a linearly polarized TCL. 
The projection shows the residual current ($J_p$) induced by the propagated laser field.
The insets illustrate the tunneling ionization of electrons at different spatial positions.
(b) The verification of the NP-CTMC (red $\blacksquare$)  by comparison with experimental results (black line) adapted from Ref.~\cite{Zhang:2020b} (see main text for details).
%
% %
Results from PC model (gray line), NP-PC (gray $\blacksquare$) and CTMC (green $\blacktriangle$) are also presented for comparison.
}
\label{fig1}
\end{figure}

We employ a comprehensive model combining nonlinear propagation and CTMC (NP-CTMC), which accounts for both the macroscopic nonlinear propagation of TCL 
and the microscopic Coulomb effect during THz generation~\cite{Chen:2026}.
Fig.\ref{fig1}(a) illustrates the model for a linearly polarized TCL.
A fundamental wave ($\omega$, 800 nm) and its second harmonic ($2\omega$, 400 nm), with an adjustable relative phase $\phi$, propagate through the gas medium and create an asymmetric field waveform that leads to ionization and associated THz generation. 
The schematic also shows electron trajectories and THz emission at different longitudinal positions. 
Simulations are performed in an axially symmetric geometry, allowing a two-dimensional representation for simplicity.
A cylindrical coordinate system is employed, % 
where $z$ is the laser propagation coordinate and $\rho$ is the radial coordinate.

The incident TCL at the entrance plane is assumed to have Gaussian envelopes in the form 
{\small
\begin{align}
\mathbf{E}_{L}(\rho, z, t) 
=& \mathrm{Re} \Bigl\{ \tilde{\mathcal{E}}_{\omega}(\rho, z, t) e^{\mathrm{i} \omega t}
+\tilde{\mathcal{E}}_{2\omega}(\rho, z, t) e^{\mathrm{i} (2\omega t + \phi)} \Bigr\}\\
\tilde{\mathcal{E}}_{j}(\rho, z, t) 
=& \frac{E_{0j} \mathcal{W}_{0j}}{\mathcal{W}_j(z)} \exp\left[-\frac{\rho^2}{\mathcal{W}_j^2(z)}\right]
\exp\left[-\mathrm{i}k_j z - \mathrm{i}k_j\frac{\rho^2}{2R_j(z)} + \mathrm{i}\zeta_j(z)\right] \nonumber \\
&\quad \times \exp\left[-2\ln 2 \frac{(t-t_0)^2}{\tau^2}\right], \qquad j = \omega, 2\omega. \nonumber
\end{align}
}

\noindent $E_{0j}$ is the peak field strength at the focus.
$\mathcal{W}_{j}(z)$ denotes the beam waist radius as a function of propagation distance, and $\mathcal{W}_{0j}$ that at the focus~\cite{Chen:2026}.
$k_{j}$ is the wavenumber,
and $\lambda_{j}$ the wavelength. 
$\zeta_j(z)$ denotes the Gouy phase acquired during propagation.
$\tau$ is the full-width-at-half-maximum (FWHM) pulse duration. $t_0$ is the pulse peak time.

The nonlinear propagation of the laser field is described by the (2D+1)-dimensional Maxwell wave equation~\cite{Esarey:1997,Takahashi:2003} 
$\nabla^{2} \mathbf{E}_{L}
-\frac{1}{c^{2}} \frac{\partial^{2} \mathbf{E}_{L}}{\partial t^{2}}$ = $\mu_{0} \frac{\partial \mathbf{J}(\rho, z, t)}{\partial t} + \frac{\omega_{0}^{2}}{c^{2}} \left(1 - \eta_{\mathrm{eff}}^{2}\right) \mathbf{E}_{L}$,
\noindent where $\omega_{0}$, $\mu_{0}$, and $c$ denote the central frequency, the vacuum permeability, and the light speed in vacuum, respectively. 
The ionization absorption loss in the source term is~\cite{Esarey:1997} %
$\mathbf{J}({\rho}, z, t)$=$\Gamma(t) n_{e}(t) I_{p} \mathbf{E}_{L}(\rho, z, t)/|\mathbf{E}_{L}(\rho, z, t)|^{2}$,
where $\Gamma(t)$, $n_{e}$, and $I_{p}$ are the ionization rate, the electron density, and the ionization potential, respectively. 
The effective refractive index, which incorporates the refraction and absorption of the neutral atom $\eta_{0}$$\sim$1, the nonlinear Kerr effect $\eta_{2} I$, and the plasma defocusing~\cite{Rae:1992}, takes the form 
$\eta_{\mathrm{eff}}=\eta_{0}+\eta_{2} I({\rho},z,t)-\omega_{p}^2/(2\omega_0^2)$.
The plasma frequency is $\omega_{p}^2$=$e^{2}n_e(t)/(\varepsilon_{0}m_{e})$. 
The electron density evolves as
$d n_{e}(t)/d t = \Gamma(t) n_{a}(t)$.
Here $n_{a}(t)$=$n_0-n_e(t)$ is the neutral density and $n_0$ the initial neutral gas density. The ionization rate $\Gamma(t)$ in tunneling regime can be estimated by the Ammosov-Delone-Krainov model~\cite{ADK_theory}.

THz generation is simulated using CTMC~\cite{Gan:2024a,Gan:2024b,Chen:2026} with the propagated TCL field.%
The ensemble-averaged acceleration follows $\langle \boldsymbol{a}(t) \rangle$ = $\sum_{i} w_{i}(t_{0}^{(i)}) \cdot \boldsymbol{a}_i(t)$, 
where $w_{i}$ and $\boldsymbol{a}_i(t)$ are the initial tunneling rate and the acceleration for the $i$-th photoelectron in the ensemble, respectively.
A Fourier transform of $\langle\boldsymbol{a}(t)\rangle$ followed by low-pass filtering to retain low-frequency components yields the THz waveform, where the THz yield is defined as its peak amplitude.
The total THz field is obtained by coherently summing the weighted local contributions~\cite{Chen:2026}.

\section{Results and Discussion}
We first perform a comparison with experimental results from Ref.~\cite{Zhang:2020b}. The experiments are performed with a linearly polarized fundamental field (800 nm, $E_{0,\omega}$~=~0.06~a.u.) and a circularly polarized second-harmonic field (400 nm, $E_{0,2\omega}$~=~$1/\sqrt{2}$$E_{0,\omega}$). This configuration is close to a linearly polarized TCL, 
thus serving as  an experimental benchmark for assessing NP-CTMC at such a laser configuration..
The NP-CTMC simulation is performed with the same parameters as the experiment for $N_2$ molecules.
The calibration of $\phi$ is performed by aligning the maximum THz yields.
Results from the PC  model, CTMC, and NP-PC are also presented for comparison.
The NP-CTMC result shows the best overall agreement with experiment.
Specifically,  the PTY properties include two main aspects, i.e., the shifts of $\phi$ for optimal (maximum) and minimum THz yields, defined as $\phi^{\mathrm{max}}$ and $\phi^{\mathrm{min}}$, respectively, and the distortion of the overall curve. 
The NP-CTMC result captures both phenomena.  

\begin{figure}[t!!]
\centering
\includegraphics[width=0.920\linewidth]{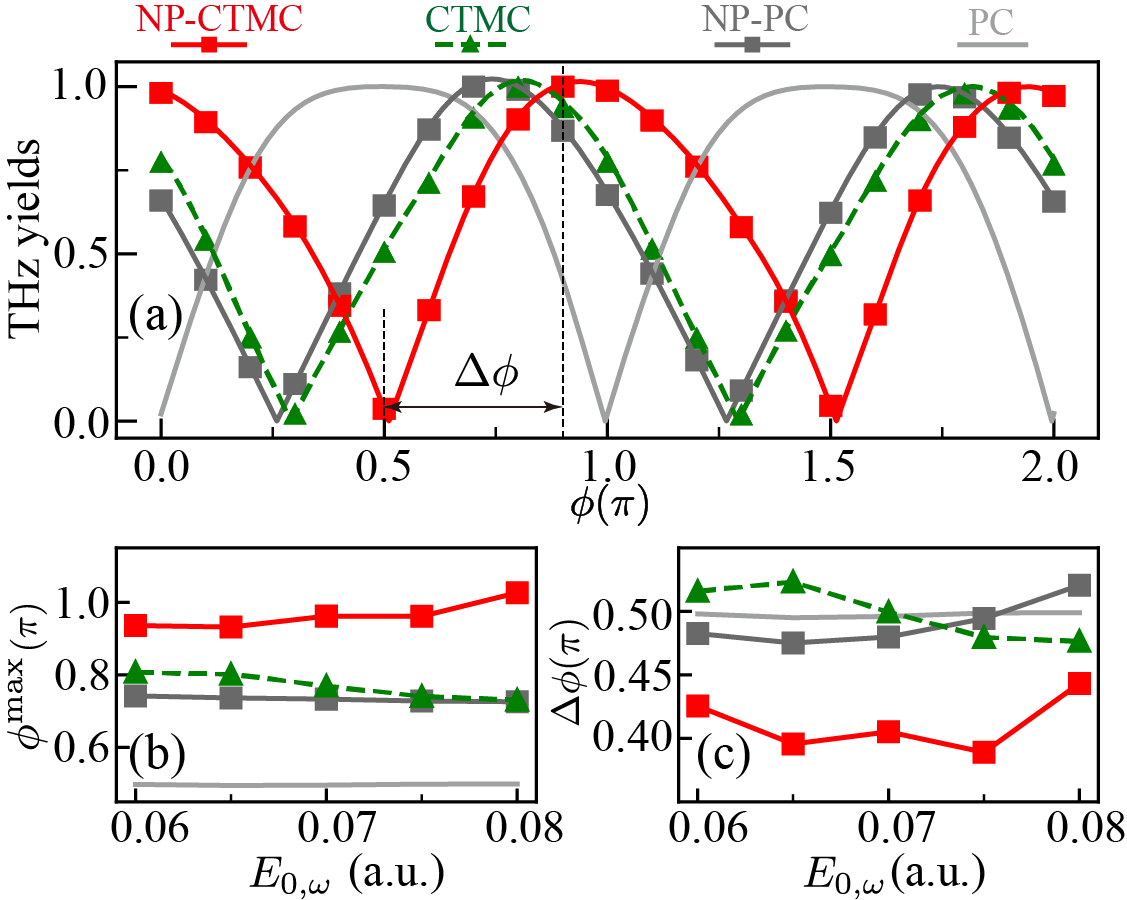}
\caption{
(a) THz yield as a function of  $\phi$ generated by a linearly polarized TCL. 
The results of the PC model (gray line), NP-PC (gray $\blacksquare$), CTMC (green $\blacktriangle$) and NP-CTMC (red $\blacksquare$) are presented for comparison. 
The field strength of the pulse is $E_{0,\omega}$=0.06 a.u. and $E_{0,2\omega}$~=~$E_{0,\omega}/4$.
Panels (b) and (c) show the corresponding $\phi^{\mathrm{max}}$ and $\Delta\phi$  at different field strengths.
}
\label{fig2}
\end{figure}

The explicit THz yields under a linearly polarized TCL are shown in Fig.\ref{fig2}(a). 
The parameters used are $E_{0,\omega}$=0.06 a.u., $E_{0,2\omega}$=$E_{0,\omega}/4$.
We present the results obtained by the PC model, NP-PC, CTMC, and NP-CTMC. 
The overall PTY profile is close to that shown in
Fig.\ref{fig1}(b). 
The PC model provides the reference properties of PTY contributed solely by the TCL without propagation effects. 
$\phi^{\mathrm{max}}$ and $\phi^{\mathrm{min}}$ are located precisely at $\phi$~=~0 and $\pi/2$, respectively,
and a symmetry about $\phi$~=~$\pi/2$ and $\pi$ is observed,
inherited from the combined TCL profile. 
By comparison with PC results, we can identify the main origin of the different properties obtained from other methods. 
The CTMC calculation includes the microscopic Coulomb effect, % 
which mainly manifests as a shift in the overall PTY curve, i.e., a shift of $\phi^{\mathrm{max}}$ and $\phi^{\mathrm{min}}$.
Similar properties are found in the NP-PC result, which includes only the nonlinear propagation effect. 
NP-CTMC includes both Coulomb and nonlinear propagation effects,  exhibiting an additional shift and distortion in the PTY curve. 
We define $\Delta \phi$=$\phi^\mathrm{{max}}-\phi^{\mathrm{min}}$ an indicator of the departure from the reference phase dependence.

Fig.\ref{fig2}(b) and (d) further show the results of $\phi^{\mathrm{max}}$ and $\Delta\phi$ at different field strengths.
Both results from the PC model remain constant since the symmetry of laser field geometry remains the same at different field strengths. 
The shifts of $\phi^{\mathrm{max}}$ induced by the Coulomb and the nonlinear propagation effects are close
and relatively stable with increasing field strength, as shown by the CTMC and NP-PC results in Fig.\ref{fig2}(b). 
However, for $\Delta\phi$, the two effects exhibit opposite trends. 
This arises because, at higher intensities, the Coulomb effect is suppressed, while the nonlinear effect is enhanced. 
NP-CTMC incorporates Coulomb-modified electron dynamics in the locally propagated laser fields.
giving its $\phi^{\mathrm{max}}$ higher than those of the other methods, and causing its $\Delta\phi$ to exhibit strong nonlinearity as a function of field strength.
%
%
% %%

\begin{figure}[ht]
\centering
\includegraphics[width=1\linewidth]{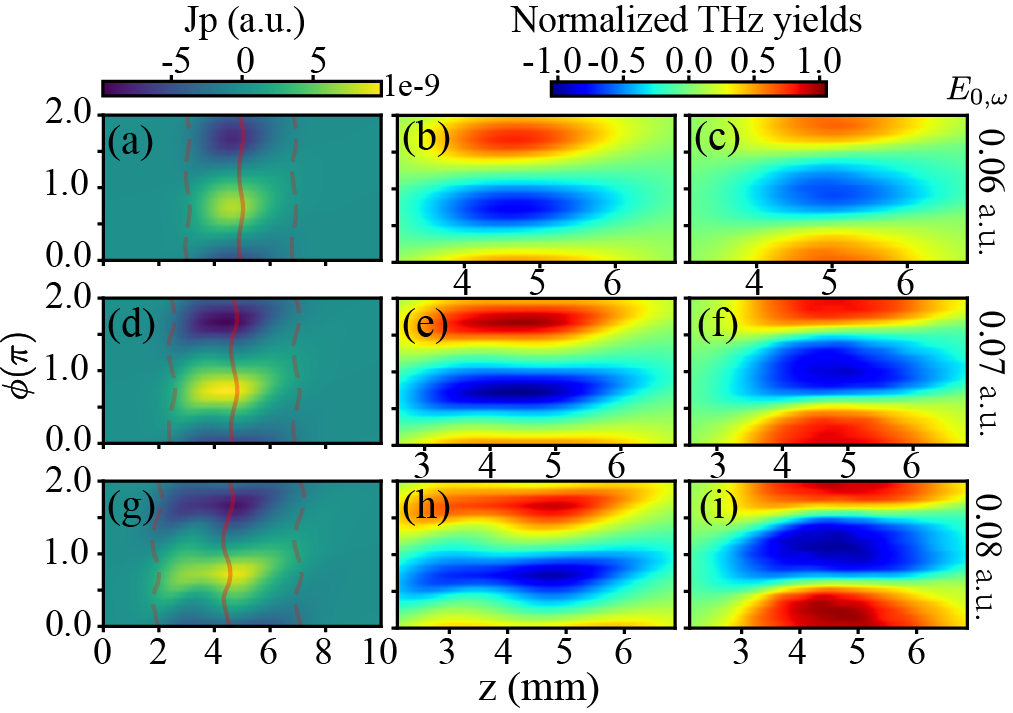}
\caption{
The left panels present the spatial distributions (along the $z$-axis) of $J_p$ as a function of $\phi$ at different field strengths as indicated on the right side. 
The middle and right panels show the corresponding normalized THz yields calculated using NP-PC and NP-CTMC, respectively.
}
\label{fig3}
\end{figure}

In Fig.\ref{fig3}, we further inspect the spatial distributions of $J_p$ in the left column and those of the THz yields calculated using NP-PC and NP-CTMC in the middle, and right columns, respectively. The rows correspond to $E_{0,\omega}$~=~0.06, 0.07, and 0.08 a.u. from top to bottom.
The $J_p$ distributions reveal the current-source structure within the filament formed by nonlinear propagation of TCL. 
The dashed lines mark the FWHM boundaries and peak positions used to define the sampling regions for NP-PC and NP-CTMC.
The THz fields emitted by these local sources interfere upon coherent superposition, contributing to the $\phi$-dependent yields shown in Fig.\ref{fig2}.
The NP-PC yield distribution directly inherits the $J_p$ profile, 
whereas NP-CTMC predicts substantial changes in the spatial distribution.
The displacement of the high-yield regions (red regions) along the $\phi$ coordinate is consistent with the shift in Fig.\ref{fig2}.
At higher field strengths, due to stronger nonlinear effects, the current-source region extends in panel (g), accompanied by a broader high-yield region in the NP-PC result in panel (h).
This broadening is less pronounced in NP-CTMC in panel (i).
These differences indicate that Coulomb-modified electron dynamics modifies the local radiating currents and thereby changes their spatial contributions to the total THz radiation~\cite{Alexandrov_2014}.

\begin{figure}[t!!]
\centering
\includegraphics[width=0.950\linewidth]{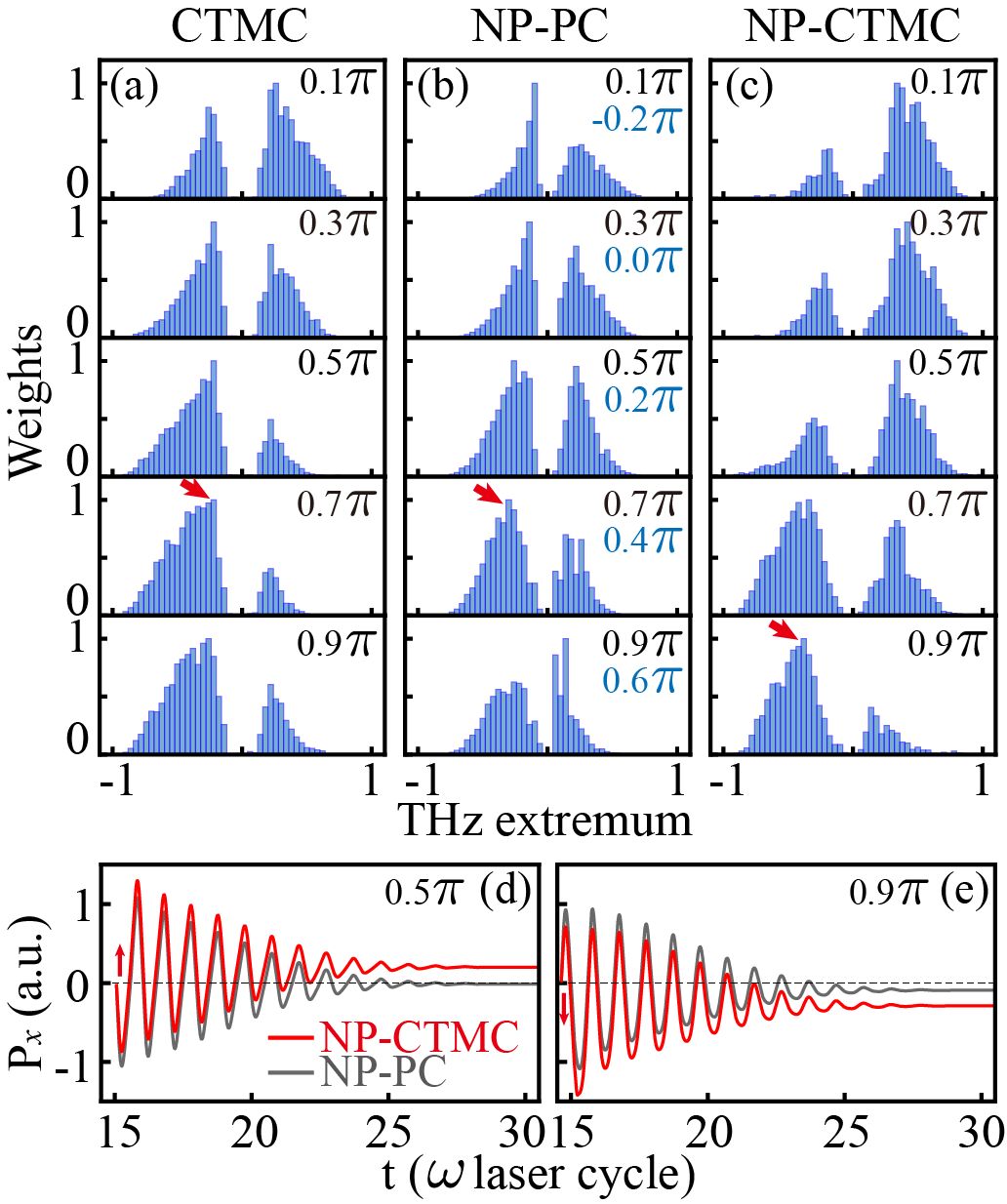}
\caption{Panels (a), (b), and (c) show normalized distributions of THz extremum radiated by electron trajectories at different $\phi$ as indicated for $E_{0,\omega}$~=~0.08 a.u., calculated via the CTMC, NP‑PC, and NP‑CTMC, respectively.
Panels (d) and (e) present momentum evolution along the polarization direction for typical trajectories from NP-PC and NP-CTMC, respectively. 
}
\label{fig4}
\end{figure}

The mechanism underlying these features can be understood by examining the electron dynamics in the nonlinearly propagated TCL. 
For the linearly polarized laser configuration, photoelectrons move predominantly along the polarization axis in either direction. 
The resulting asymmetry in electron motion contributes to the net current responsible for THz emission. 
According to the well-established recollision model~\cite{Liu:2014}, the final electron emission direction is associated with the initial tunneling time and the subsequent ionization dynamics governed by the combined effects of Coulomb potential and TCL. 
Fig.\ref{fig4} shows the normalized distributions of THz extrema radiated by individual electron trajectories at $E_{0,\omega}$=0.08 a.u. 
Results for different $\phi$ are shown for CTMC, NP-PC, and NP-CTMC calculations as indicated. 
The asymmetry is largest near $\phi$=0.7$\pi$ for CTMC and NP-PC, whereas it reaches its maximum near $\phi$=0.9$\pi$ for NP-CTMC, consistent with $\phi^{\mathrm{max}}$ in Fig.\ref{fig2}(a).

NP-CTMC can be essentially regarded as the coherent sum of a series of CTMC calculations with a distorted laser field. 
The panels in (b) show the approximate effective $\phi$ (blue text) induced by the nonlinear propagation effect. 
By connecting the the fourth (or fifth) panel in (b) with the second and third panels (or the third and fourth panels) in (a), which contain the asymmetric emission produced mainly by recollision dynamics under a standard laser profile, 
we can qualitatively infer the relevant asymmetry distribution observed in (c).  
Moreover, a signature unique to the NP-PC and NP-CTMC results, with the latter being more pronounced, is the increase in the peak position of the distribution as indicated by the red arrows. This arises from the higher asymmetry in the distorted TCL profile caused by nonlinear propagation, which is further enhanced by the Coulomb effect in NP-CTMC.
This can be illustrated directly by the momentum evolution of  typical trajectories shown in panels (d) and (e) for NP-PC and NP-CTMC, at  $\phi=0.5\pi$ and $0.9\pi$, respectively.
The electron release time is near the peak position of the laser pulse, while the transverse momentum is assumed to be zero. In Panel (d), Coulomb effect enhances recollision, leading to a change in final emission direction from negative (NP-PC) to positive (NP-CTMC). In contrast, for $\phi=0.9\pi$, the recollision causes a further shift of the emission direction toward the negative direction. The main recollision is roughly indicated by red arrows in the panels.

In summary, we have investigated the $\phi$-dependent THz yield in a linearly polarized TCL plasma, using a model that combines nonlinear propagation with microscopic Coulomb effects. 
Nonlinear propagation distorts the local TCL waveform, while the Coulomb interaction further modify the electron dynamics driven by this field, which affects the asymmetry of electron motion along the polarization axis and the resulting THz radiation.
Compared with models that account for either effect alone, the combined model predicts changes in the spatial distribution of THz sources, an additional shift of the optimal phase $\phi^{max}$, and a distortion of the PTY curve.
These results provide a basis for investigating THz optimization in more complex configurations, such as cascaded multicolor laser schemes~\cite{Martinez2015_multi_color_THz}.

\smallskip
\begin{backmatter}
\bmsection{Funding} This work was supported by  the National Natural Science Foundation of China (No. 12541505, No. 12334011, No. 12674347, No. 12374262, and No. 12674346),  the Natural Science Foundation of Henan (Grant No. 252300421304), the National Key Research and Development Program of China (No. 2022YFA1604302), and the Natural Science Foundation of Shanghai (No. 25ZR1401248). 
%\smallskip
\bmsection{Acknowledgements}We acknowledge the computational resource of the HPC platform in ShanghaiTech University.
%\smallskip
\bmsection{Disclosures} The authors declare no conflicts of interest.
%\smallskip
\bmsection{Data availability} Data underlying the results presented in this paper are not publicly available at this time but may be obtained from the authors upon reasonable request.
\end{backmatter}

	% Bibliography
	\bibliography{main}

	\bibliographyfullrefs{main}
	
	%
	%%%%%%%%%%%%%%%%%%%%%%%%%%%%%%%%%%%%%%%%%%%%%%%%%%%%%%%%%%%%%%%%%%
	% 

\end{document}